\documentclass[%
 reprint,
superscriptaddress,
 amsmath,amssymb,
 aps,
 prb,
showpacs
]{revtex4-2}

\usepackage{color}

\usepackage{graphicx, float}
\usepackage{dcolumn}
\usepackage{bm}
\usepackage{comment}
\usepackage{mathtools}
\DeclareMathOperator*{\Motimes}{\text{\raisebox{0.25ex}{\scalebox{0.8}{$\bigotimes$}}}}

\usepackage{color}

\begin{document}

\preprint{APS/123-QED}

\title{Impact of planar defects on the reversal time of single magnetic domain nanoparticles}

\author{Hugo Bocquet}
\email{h.bocquet@protonmail.ch}
\affiliation{Laboratory for Theoretical and Computational Physics, Paul Scherrer Institut, CH-5232 Villigen PSI, Switzerland}
\author{Armin Kleibert}
\affiliation{Swiss Light Source, Paul Scherrer Institut, CH-5232 Villigen PSI, Switzerland}
\author{Peter M. Derlet}
\email{peter.derlet@psi.ch}
\affiliation{Laboratory for Theoretical and Computational Physics, Paul Scherrer Institut, CH-5232 Villigen PSI, Switzerland}

\date{\today}

\begin{abstract}  
Recent experimental investigations of individual magnetic nanoparticles reveal a diverse range of magnetic relaxation times which cannot be explained by considering their size, shape, and surface anisotropy, suggesting other factors associated with the internal microstructure of the particles are at play. In this letter, we apply Langer's theory of thermal activation to single magnetic domain fcc Co nanoparticles, whose experimental microstructures are characterized by planar defects, and derive an analytic expression for the relaxation time. The obtained Arrhenius exponential and its prefactor, which is often assumed to be a constant, are here found to both depend exponentially on system size and the number of defects. Together they provide a quantitative prediction of the experimental findings, and more generally highlight the importance of structural defects when considering magnetic stability. 
\end{abstract}

\maketitle

Magnetic nanoparticles exhibit unique properties such as single domain states, superparamagnetism and enhanced magnetic moments, making them technologically interesting in fields ranging from spintronics, information and energy storage, and bio-engineering applications~\cite{BCH94,Kodama99,BBBB+05,FPCS09,Jones11}. However, despite impressive progress in their synthesis, the resulting magnetic properties often deviate from expectations based on commonly applied scaling laws~\cite{SMWF+00,LSS07,AGS+11}. These discrepancies hamper the use of magnetic nanoparticles in practical applications and question our fundamental understanding of nanoparticle magnetism. This is partly due to the difficulties of experimental characterization, where measurements often integrate over large particle ensembles with varying particle sizes and microstructures, and including complex interactions~\cite{MHF10,KFOG+11,VBKH+13}. One aspect not yet explicitly considered is that such nanoparticles often contain structural defects and therefore a microstructure that has no counterpart in the respective bulk~\cite{Marks94,BF05,CZWC+13,YCSO+17}. Hence, simple scaling laws extrapolating bulk properties to the nanoscale often fail to achieve a realistic description of the magnetic properties such as the magnetic relaxation time of nanoparticles even in the case of pure $3d$ transition metals~\cite{Kleibert2014, Kleibert2017}. 

In this Letter, we demonstrate that the presence of planar defects, such as twin boundaries and stacking faults, can modify the magnetic reversal time (the relaxation time for reversing the magnetization) of single magnetic domain fcc Co nanoparticles by several orders of magnitude. By using the formalism of Langer~\cite{Langer1969} to evaluate the activation rate (the inverse of the reversal time), we can account for magnetic heterogeneities coming from the structural defects and extend the seminal work of Brown on the magnetic relaxation of nanoparticles~\cite{Brown63}. By exploiting the long wavelength limit of the anisotropy field, we derive an explicit equation for the prefactor in which the usual entropic contribution~\cite{Desplat2018, Desplat2020} is expressed in terms of spin waves, giving an exponential dependence in particle diameter and the number of planar defects. Whilst conventional wisdom cautions application of Langer's approach to low damping coefficient materials~\cite{Coffey2012}, recent work by two of the present authors demonstrate that Langer's approach can remain valid for the small damping coefficient of Co, when the system satisfies an appropriate equilibrium condition~\cite{prb}. Combined, these developments allow us to justify unambiguously experimental findings in size-selected, individual Co nanoparticles~\cite{Kleibert2017}. In general, we see that the enhanced anisotropy energy due to the reduced symmetry of the structural defects has a much stronger effect on the magnetic stability of a nanoparticle than, for example, the non-colinear spin structure at the surface induced by the lower coordination~\cite{JWTM+01,GK03}.

\begin{figure}[h]
\includegraphics[width=0.85\linewidth,trim=0.1cm 0.4cm 0.01cm 0cm, clip]{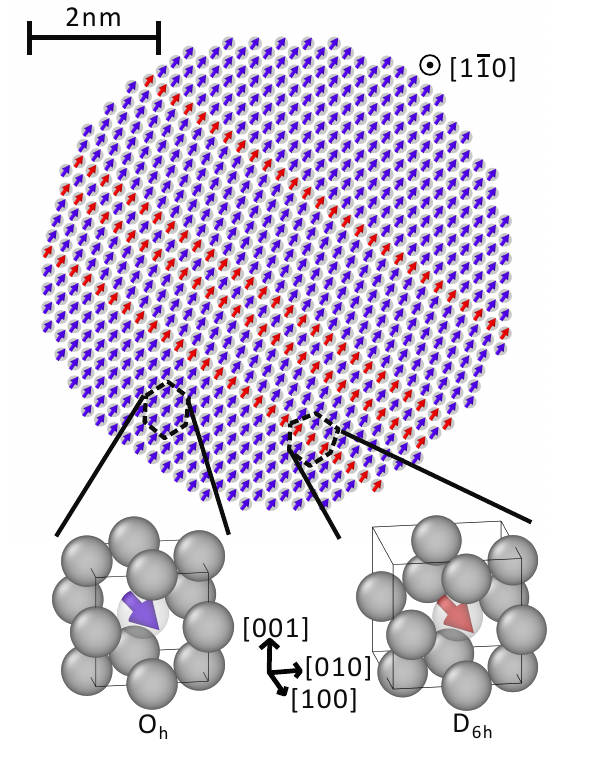}
\caption{\label{fig:model} Atomistic model of a spherical Co nanoparticle with a diameter of 6~nm viewed along the $\left[1\overline{1}0\right]$-direction of the fcc lattice. Planar defects across the particle are identified by the reflection of the crystal orientation and the corresponding spins are highlighted in red. Enlarged views on the two different types of nearest neighbour environments are shown at the bottom. The fcc-sites exhibit octahedral $O_h$-symmetry, while the defective sites within the planar defects exhibit dihedral $D_{6h}$-symmetry supporting therefore different magnetic anisotropy terms.}
\end{figure}

A general treatment of the role of defects on the magnetic properties of nanoparticles is complex, as it requires one to develop a microscopic description of the structure and its effect on the relevant magnetic degrees of freedom. This can be however achieved effectively by considering magnetic moments localized at the atomic sites embedded in site-dependent crystal fields. For the present work we consider fcc Co nanoparticles with planar defects such as twin boundaries and stacking faults. Such defects are frequently found experimentally and occur also in other magnetic fcc metal nanoparticles~\cite{SKSS+97,DRGG01,YALG+01,Wernsdorder2002,TSEK+13,Kleibert2017,VSBL+23}. A respective nanoparticle model is shown in Fig.~\ref{fig:model}. At the atomic scale, the sites composing the planar defects are in a dihedral $D_{6h}$ nearest neighbour environment, similar to the sites in the equivalent hcp-lattice, while the sites belonging to the regular fcc-lattice are in an octahedral $O_h$ environment as shown in the insets in Fig.~\ref{fig:model}. Correspondingly, magnetic moments at the dihedral $D_{6h}$ sites exhibit a uniaxial anisotropy, while, in contrast, the octahedral $O_h$-sites allow at most for a leading order 8-fold symmetric magnetic anisotropy terms. In Co the magnetic moment per spin is practically identical in both symmetries, while the single ion magnetic anisotropy differs significantly~\cite{Ono1990, Kleibert2017}. Hence, we can write the spin Hamiltonian for such a particle in the basis of the  fcc lattice vectors:
\begin{equation}
\begin{split}\label{eq:hamiltonian}
    \mathcal{H}=&-\frac{J}{2}\sum_{\left<i,j\right>}\bm{s}_i \cdot \bm{s}_{j} \\&- K_{O_h} \sum_{i \in O_h} (s_{i}^x)^2(s_{i}^y)^2 
    + (s_{i}^x)^2(s_{i}^z)^2 + (s_{i}^y)^2(s_{i}^z)^2 \\&- \frac{2}{3}K_{D_{6h}} \sum_{i \in D_{6h}} (s_{i}^x)(s_{i}^y) + (s_{i}^x)(s_{i}^z) + (s_{i}^y)(s_{i}^z)\;,
\end{split}
\end{equation}
with unit spins, the exchange $J$ and the site-dependent anisotropy constants $K_{O_h}$ and $K_{D_{6h}}$. The exchange interaction takes into account the twelve nearest neighbours and is assumed to be unaffected by the lower symmetry of the defects. 

\begin{figure}[h]
\includegraphics[width=0.85\linewidth, trim=0cm 0.5cm 0cm 0.8cm, clip]{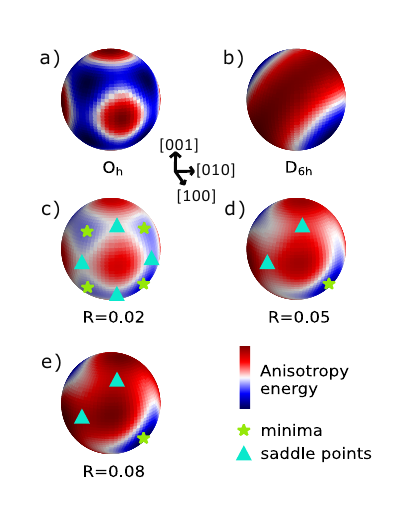}
\caption{\label{fig:anisotropyEnergy} Magnetic anisotropy energy for (a) the $O_h$ symmetry of the fcc lattice and (b) the $D_{6h}$ symmetry induced by the [111]-oriented stacking faults of the fcc-lattice. (c-e) Magnetic anisotropy energy for nanoparticles obtained by insertion of a number of stacking faults, which corresponds to a fraction of defective $D_{6h}$ sites given by the defect ratio $R$ as defined in the text. The minima and saddle points are highlighted with stars and triangles, respectively. The 8-fold degeneracy of the fcc-particle is lifted into a 2-fold degeneracy by the presence of the defects. For $R>0.03$, the anisotropy energy exhibits only two minima connected by six saddle points.} 
\end{figure}

Using $J=37$~meV and the bulk anisotropy for the fcc and hcp Co, respectively $K_{O_h}=5$~$\mu$eV and $K_{D_{6h}}=35$~$\mu$eV~\cite{Ono1990, Kleibert2017}, we first identify the ground state configurations, as being a pair of uniaxial ferromagnetic configurations along the easy axis anisotropy of the fcc latice which coincides with one the [111]-oriented easy axes of the defects, as represented in Fig.~\ref{fig:anisotropyEnergy}(a) and~(b). We then verify with the help of mART~\cite{Bocquet2023}, which searches for saddle-points within the magnetic energy landscape, that Co particles up to 33~nm remain in a single ferromagnetic domain state when reorientating (see appendix~\ref{sec:uniformTransition}).

This result allows us to describe the magnetic anisotropy energy of the nanoparticle as being only a function of the magnetization orientation. Therefore, Fig.~\ref{fig:anisotropyEnergy}(c--e) shows the magnetic anisotropy energy for a nanoparticle as obtained by the insertion of [111]-oriented stacking faults to the defect-free fcc-particle, such that $R$ is the fraction of atoms in the planar defects (see red spins in Fig.~\ref{fig:model}) with respect to the total number of atoms. Inspection of the stability of the states reveals that the numerous (meta-)stable states of the fcc-particle are simplified to the pair of uniaxial ground states when $R>0.03$, which corresponds to only one stacking fault in the middle of a 20-nm-particle. In this regime, the ground states are connected by one of six equivalent saddle points, that support the thermally activated transition. This defect-driven reduction to a uniaxial symmetry fundamentally explains the experimental findings by Wernsdorfer \textit{et al.}~\cite{JWTM+01,Wernsdorder2002}. 

 Although the presence of defects in the Co nanoparticle simplifies the problem down to a unique type of thermally activated transition corresponding to a uniform flip of the magnetization, the magnetic reversal time depends strongly on the number of defects and the size of the particle. We will see this dependence by evaluating the activation rate, $\Gamma$, obtained by computing the flux of the probability density at the saddle point for a system which is assumed to be in equilibrium at the minimum~\cite{Langer1969}. The flux is calculated taking into account thermal fluctuations according to stochastic Landau-Lifschitz dynamics~\cite{prb} --- an aspect not considered in Refs.~\cite{Bessarab2012,Bessarab2013}, where only the precessional flux is evaluated. 
The central result is an Arrhenius law,
\begin{equation}\label{eq:transitionRate}
   \Gamma=  \nu e^{-\Delta E/K_BT}\;,
\end{equation}
with the prefactor,
\begin{equation}\label{eq:prefactor}
\nu=6\frac{|\kappa|}{2\pi \hbar} \sqrt{\prod_{n=1}^{2N}\frac{ \varepsilon^{(m)}_{n}}{|\varepsilon^{(s)}_{n}|}}\;.
\end{equation}
Here, the prefactor contains the growth rate $\kappa$ of the unstable mode of the linearized dynamics at the saddle point, i.e. the speed at which the magnetization goes away from the saddle point (see appendix~\ref{sec:transitionRate}). 
The energies $\varepsilon^{(m)}_n$, $\varepsilon^{(s)}_n$ are the eigenvalues of the Hessian of the energy at the minimum and at the saddle point. The ratio of their product at the minimum and at the saddle point can be interpreted as an entropic contribution coming from the difference in number of thermally accessible states~\cite{Desplat2018, Desplat2020, prb}. The Hessian is computed on the transverse planes to the spins as $ \partial_j^\beta \partial_i^\alpha \mathcal{H}-\delta_{ij} \delta^{\alpha \beta}\sum_{\gamma=1}^{3}  s_i^\gamma \partial_i^\gamma \mathcal{H}$ due to the curved spin space originating from the fixed spin magnitude. The factor of 6 at the front accounts for the six equivalent transitions in our problem.

Numerically evaluating the prefactor, Eq.~(\ref{eq:prefactor}), for the experimentally relevant particle size range is computationally expensive, since it requires the diagonalization of a $2N\times2N$ Hessian, where $N$ can exceed $10^6$. However, if we neglect boundary effects and take the long wavelength limit of the Fourier transform of the anisotropy as suggested by the spatial extent of the defects, the Hamiltonian becomes translationally invariant and the dynamical excitations feature spin waves, whose frequencies write directly in terms of the eigenvalues of the Hessian~\cite{prb}. 

With these simplifications, the prefactor becomes
\begin{equation}\label{eq:prefactorSpinWaves}
    \nu=6\frac{w_{\bm{0}}^{(m)}}{2\pi}\prod_{\bm{q}\neq \bm{0}} \frac{w_{\bm{q}}^{(m)}}{w_{\bm{q}}^{(s)}}\;,
\end{equation}
where $w_{\bm{q}}^{(m)}=\sqrt{\varepsilon_{\bm{q},1}^{(m)}\varepsilon_{\bm{q},2}^{(m)}}$ and $w_{\bm{q}}^{(s)}=\sqrt{\varepsilon_{\bm{q},1}^{(m)}\varepsilon_{\bm{q},2}^{(m)}}$ are the degenerate spin wave frequencies at the minimum and at the saddle point respectively obtained for the small damping factor $\alpha=10^{-2}$ of Co~\cite{Weber2019} (see appendix~\ref{sec:longWaveLength}). Note that in this regime, the growth rate $\kappa$ corresponding to the instability along which the transition proceeds to the saddle point, at $\bm{q}=\bm{0}$, no longer appears. 

\begin{figure}[h]
\includegraphics[width=0.9\linewidth,trim=0.2cm 0.3cm 0cm 0cm, clip]{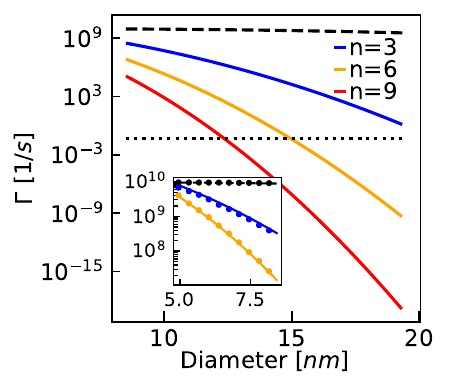}
\caption{\label{fig:transitionRate} Activation rate for a Co particle as a function of the diameter for different numbers of stacking faults or pairs of twin boundaries $n$ according to Eqs.~(\ref{eq:prefactorAnaly}) and~(\ref{eq:arrheniusAnaly}). Particles with an activation rate smaller than $1/20s$ (dotted black line) present a stable magnetic contrast in the experiment of Ref.~\cite{Kleibert2017}. The dashed line represents the $R\rightarrow 0$ limit of Eqs.~(\ref{eq:prefactorAnaly}) and~(\ref{eq:arrheniusAnaly}), corresponding to a defect-free particle. In the inset with the same axis labels, the result is compared to the the activation rate evaluated numerically (dots) with the prefactor from Eq.~(\ref{eq:prefactor}).}
\end{figure}

To evaluate the ratio of the product of the spin wave frequencies, representing the original entropic contribution, we take advantage of the fact that the spin wave spectra are dominated by the large exchange contribution in the Hamiltonian ($J \gg K_{O_h}, K_{D_{6h}}$) and are therefore constantly gapped by the mean curvature of the anisotropy energy: for instance at the minimum, $\hbar w^{(m)}_{\bm{q}}=\hbar w_{\bm{q}}+\mathcal{C}^{(m)}$ with $\mathcal{C}^{(m)}$ the mean curvature of the anisotropy at the minimum. The curvatures can be observed in Fig.~\ref{fig:anisotropyEnergy} by considering the gradient of colors. The prefactor, Eq.~(\ref{eq:prefactorSpinWaves}), depends thus exponentially on the mean curvatures at the minimum and at the saddle point, which in turn depends on the defect fraction $R$. For $R\ll 1$, we have (see appendix~\ref{sec:longWaveLength}):
\begin{multline}\label{eq:prefactorAnaly}
    \nu=\frac{3}{\pi \hbar}\left(\frac{4}{3}K_{O_h}+2RK_{D_{6h}}\right) \\
        \times \exp\left[\frac{N}{17J}\left(\frac{5}{3}K_{O_h}+6RK_{D_{6h}}\right)\right]\;.
\end{multline}
The term at the front in parentheses is the mean curvature of the anisotropy at the minimum $\mathcal{C}^{(m)}$ coming from $w_{\bm{0}}^{(m)}$ and the term in parantheses in the exponential contains the difference between the mean curvature at the minimum and at the saddle point. Also for $R\ll 1$, we can write
\begin{equation}\label{eq:arrheniusAnaly}
   e^{-\Delta E/K_BT}= \exp\left[-\frac{N}{K_BT}\left(\frac{K_{O_h}}{12}+RK_{D_{6h}}\right)\right]\:.
\end{equation}
The activation rate as obtained from Eqs.~(\ref{eq:prefactorAnaly}) and~(\ref{eq:arrheniusAnaly}) is plotted in Fig.~\ref{fig:transitionRate} for different numbers of typical twin boundaries and stacking faults in the Co particle instead of $R$. These defects are typical in the sense that they are assigned a value of $R$ corresponding to the mean cross section $\pi^2 d^2/64$ when they are uniformly distributed along the particle diameter $d$.

The simplification associated with the long wavelength limit of the anisotropy field can be justified by comparing to the exact numerical evaluation of Eq.~(\ref{eq:prefactor}) for small model particles with randomly distributed planar defects reflecting a particular value of $R$ --- see inset of Fig.~\ref{fig:transitionRate} which shows very good agreement with the analytical predictions. More generally, the validity of the magnetic activation rate calculation depends on the assumption of equilibrium at the minimum, which can be assessed by comparing the time to reach equilibrium in the minimum basin, $1/2\alpha w_{\bm{0}}^{(m)}$, with the inverse activation rate~\cite{prb}. For the aforementioned damping factor $\alpha$, we validate the calculation for particles bigger than  8~nm containing more than two stacking faults, as these have time to reach equilibrium between reversal events. This is due to the fact that the longest precessional time, $1/w_{\bm{0}}^{(m)} = \hbar/\mathcal{C}^{(m)}$, decreases with the number of defects. 

In contrast to previous estimates of the transition rate of a nanoparticle~\cite{Kleibert2017},  the obtained equation for the prefactor, Eq.~(\ref{eq:prefactorAnaly}), has a GHz factor arising from the anisotropy gap and a factor which depends exponentially on system size and the number of defects. This last factor can vary between $\sim 1$ for a particle of 8~nm with 3 stacking faults ($n=3$) to $\sim 10^{11}$ for a particle of 20~nm with 9 stacking faults ($n=9$). Such a dramatic change of the prefactor compensates partially the drop in the Arrhenius exponential as the system size and the number of defects increase. However, this effect is not described by the well-known Meyer-Neldel compensation rule~\cite{Yelon1990, Yelon1992}, as the energy barrier cannot be factored out from the exponent in the prefactor. Nonetheless, we conclude from Eqs.~(\ref{eq:prefactorAnaly}) and~(\ref{eq:arrheniusAnaly}) that the compensation is limited, because the energy of the fastest spin wave excitation ($\approx17J$) is too large compared to the thermal energy scale $K_BT$ at room temperature, suggesting that compensation depends on the thermal accessibility to the spin wave spectrum. As a consequence, and as observed in Fig.~\ref{fig:transitionRate}, the activation rate $\Gamma$ decreases with respect to both the particle diameter and defect density. 

A small number of defects can therefore modify the activation rate by many orders of magnitude, an effect that is significantly stronger than that due to the surface anisotropy~\cite{GK03,YCKG+07} and comparable to the size effect. In particular, we see in Fig.~\ref{fig:transitionRate} that doubling the number of stacking faults has an impact comparable to increasing the diameter of the particle by 50\%. This new microstructural dependence explains the behaviours seen in experiment, where Co nanoparticles smaller than 20~nm can exhibit stable magnetic states at room temperature. For Co particles with a diameter of about 15~nm or larger, our calculations show that the presence of at least six stacking faults, or equivalently twelve twin boundaries, are sufficient to yield an activation time larger than 20~sec, which is the experimental resolution time distinguishing superparamagnetism from stable magnetism in Ref.~\cite{Kleibert2014, Kleibert2017}. 

In conclusion, we have demonstrated that the presence of planar defects in fcc Co magnetic nanoparticles can fundamentally affect its low energy thermally activated transition mechanisms, resulting in several orders of magnitude increase in the timescale of magnetic reversal and more generally the superparamagnetic fluctuation timescale. The analytical approach developed in this letter paves the way for gaining quantitative insight into the relaxation time of nanomagnetic systems containing internal structural heterogeneities, allowing for an investigation of magnetic stability and tunability at the nanoscale as function of microstructure.

\begin{acknowledgments}
The present work was supported by the Swiss National
Science Foundation under Grant No. 200021-196970.
\end{acknowledgments}
\appendix

\section{Single-domain ferromagnetic states for Co nanoparticles}\label{sec:uniformTransition}
For the Hamiltonian of Eq.~(1), we want to check whether low energy landscape between the uniaxial ground states of the Co nanoparticle $\bm{s}_i=\pm(1,1,1)^T/\sqrt{3}$ is composed of single-domain ferromagnetic states. We can treat the worst case when all the sites have $D_{6h}$ symmetry and therefore exhibit the largest anisotropy $K_{D_{6h}}$ which minimizes the exchange length $\sim \sqrt{J/K}$. This corresponds to the problem investigated in Ref.~\cite{Bocquet2023}, as the ideal hcp and fcc lattices have an equivalent nearest-neighbour bonding network. For this problem, it turns out that a non-uniform transition state starts to appear for a critical diameter equal to $4.0\sqrt{J/K_{D_{6h}}}a_{HCP}=2.9\sqrt{J/K_{D_{6h}}}a_{FCC}$, where $a_{HCP}$ and $a_{FCC}$ are the respective lattice parameters. For $a_{FCC}=0.36nm$, the critical diameter is 33nm. In other words, we can conclude that for Co particles smaller than 33nm the exchange length is too large compared to the particle size to yield a non-uniform transition.

\section{Computing the transition rate}\label{sec:transitionRate}
Following the result and the terminology of Ref.~\cite{prb}, we want to evaluate the transition rate. The latter writes as in Eqs.~(2) and~(3) in the main manuscript. To compute the eigenvalues of the Hessian, we first evaluate it in the $3N$ space, giving
\begin{equation}\label{eq:Hessian}
\mathrm{H}_{ij}^{\alpha \beta}= \partial_j^\beta \partial_i^\alpha \mathcal{H}-\delta_{ij} \delta^{\alpha \beta}\sum_{\gamma=1}^{3}  s_i^\gamma \partial_i^\gamma \mathcal{H} \;,
\end{equation}
where $\alpha$, $\beta$, $\gamma$ are regular spatial coordinates and $\mathcal{H}$ is the Hamiltonian of Eq.~(1) in the main manuscript, such that we can write explicitly for any state $\{\bm{s}_i\}$:
\begin{equation}
\begin{split}
    \mathrm{H}^{\alpha \beta}_{ij} =& J \delta^{\alpha \beta} \left(z \delta_{ij}-\delta_{\left<i,j\right>}\right)\\
    &+2K_{O_h} \delta_{ij} \delta_{i\in O_h} \delta^{\alpha \beta} \left((s_i^\alpha)^2-\sum_\gamma (s_i^\gamma)^4 \right)\\
    &+\frac{2}{3} K_{D_{6h}} \delta_{ij} \delta_{i\in D_{6h}} \left(\delta^{\alpha \beta} \sum_{\gamma \delta}  s_i^{\gamma} s_i^{\delta} - 1 \right)\;,
\end{split}
\end{equation} 
where $z=12$ is the number of nearest neighbours. The $2N$ physical eigenvalues for the calculation of the prefactor are obtained by diagonalizing the Hessian on the tangent space, i.e. the space transverse to the spins described by the vector $\bm{\eta}$ ($T_{\{\bm{s}_i\}}\mathcal{M}=\{\bm{\eta}|\bm{\eta}_i \cdot \bm{s}_i=0 \forall i\}$  with $\bm{\eta}=\Motimes_i \bm{\eta}_i$), at the stationary state $\{\bm{s}_i\}$ (minimum or saddle point). Also the harmonic expansion about these states writes in the tangent space: 
\begin{equation}
\mathcal{H}^{(m)}(\bm{\eta})= E^{\{\bm{s}_i\}} + \frac{1}{2} \bm{\eta}^T\mathrm{H}^{(m)} \bm{\eta}\;,
\end{equation}
\begin{equation}
\mathcal{H}^{(s)}(\bm{\zeta})= E^{(s)} + \frac{1}{2} \bm{\zeta}^T\mathrm{H}^{(s)} \bm{\zeta}\;,
\end{equation}
with $\bm{\zeta}$ the vector describing the tangent space at the saddle point to distinguish from the minimum. The dynamical mode equation from which the growth rate $\kappa$ is derived writes
\begin{equation}\label{eq:dynamicalMode_min}
    i \hbar w^{(m)} \bm{\eta}_{w}=\left(\mathrm{U}+\alpha\mathrm{I}\right)\mathrm{H}^{(m)}  \bm{\eta}_{w}\;,
\end{equation}
\begin{equation}\label{eq:dynamicalMode_sp}
    i \hbar w^{(s)} \bm{\zeta}_{w}=\left(\mathrm{U}+\alpha\mathrm{I}\right)\mathrm{H}^{(s)}  \bm{\zeta}_{w}\;,
\end{equation}
where $\bm{\eta}_{w}=\int \bm{\eta} e^{i wt}dt$ (similar for $\bm{\zeta}_w$) and $\hbar$ arises actually from $S(1+\alpha^2)/\gamma$ where $S$ is the unit spin moment magnitude and we consider $\alpha \ll 1$. $\mathrm{I}$ is the identity matrix coming from the dissipative part of the dynamics ($\alpha=0.01$ is the damping factor for Cobalt~\cite{Weber2019}) and $\mathrm{U}$ is an orthogonal matrix coming from the precessional part of the dynamics and which writes with the diagonal blocks:
\begin{equation}\label{eq:dynamicTensor}
    \left(\mathrm{U}\right)_i=
    \begin{pmatrix}
        0 & 1 \\ -1 & 0 
    \end{pmatrix}
    \;.
\end{equation}
The only real negative solution $i\hbar w^{(s)}$ at the saddle point in Eq.~(\ref{eq:dynamicalMode_sp}) corresponds to the growth rate of the unstable mode $\kappa$.

\section{Expressing the transition rate in the long-wavelength limit of the anisotropy field}\label{sec:longWaveLength}
\begin{figure}[h]
\includegraphics[width=1.0\linewidth,trim=0cm 0cm 0cm 0cm, clip]{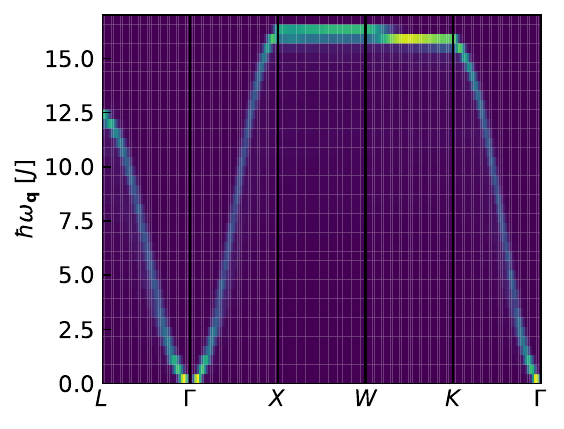}
\caption{\label{fig:spectrum} Exchange spin wave spectrum for a 5nm particle along the principal directions of the FCC lattice. Typical for a ferromagnet, there is a quadratic dispersion at the $\Gamma$ point. The rest of the spectrum is typical for spin waves on an FCC lattice~\cite{Chimata2017}. }
\end{figure}

We consider the Hessian at the uniform stationary state $\{\bm{s}_i\}$ in reciprocal space ($\mathrm{H}_{\bm{qq}'}^{\alpha \beta}=\sum_{ij}\mathrm{H}_{ij}^{\alpha \beta} e^{i(\bm{q}\bm{r}_i+\bm{q}'\bm{r}_j)}$):
\begin{equation}
\begin{split}
    \mathrm{H}^{\alpha \beta}_{\bm{qq}'}=&2 J\delta(\bm{q}+\bm{q}')\delta^{\alpha \beta} \sum_{\bm{a}:|\bm{a}|>\bm{0}}\left[1-\cos(\bm{q}\bm{a})\right]\\
    &+2 K_{O_h}(\bm{q}+\bm{q}')  \delta^{\alpha \beta} \left((s_i^\alpha)^2-\sum_\gamma (s_i^\gamma)^4 \right)\\
    &+\frac{2}{3} K_{D_{6h}}(\bm{q}+\bm{q}') \left(\delta^{\alpha \beta} \sum_{\gamma \delta}  s_i^{\gamma} s_i^{\delta} - 1 \right)\;.
\end{split}
\end{equation}
The Fourier components of the anisotropy fields are given by
\begin{equation}
    K_{O_h}(\bm{q})=K_{O_h}\sum_{i \in O_h} e^{i\bm{q}\bm{r}_i}\;,
\end{equation}
and
\begin{equation}
    K_{D_{6h}}(\bm{q})=K_{D_{6h}}\sum_{i \in D_{6h}} e^{i\bm{q}\bm{r}_i}\;.
\end{equation}
Because of the spatial extent of the defects, which constitute a cross section of the particle, we consider the long-wavelength limit of the anisotropy fields, i.e. $\bm{q} \rightarrow \bm{0}$. It follows that $K_{O_h}(\bm{q})=\delta_{\bm{q}}K_{O_h}$ and $K_{D_{6h}}(\bm{q})=\delta_{\bm{q}}RK_{D_3h}$ at leading order in $R\ll 1$, the fraction of $D_{6h}$ sites. The spatial degrees of freedom in the Hessian are now diagonal in reciprocal space:
\begin{equation}
\begin{split}
    \mathrm{H}^{\alpha \beta}_{\bm{q}}=&2 J\delta^{\alpha \beta} \sum_{\bm{a}:|\bm{a}|>\bm{0}}\left[1-\cos(\bm{q}\bm{a})\right]\\
    &+2 K_{O_h} \delta^{\alpha \beta} \left((s_i^\alpha)^2-\sum_\gamma (s_i^\gamma)^4 \right)\\
    &+\frac{2}{3} RK_{D_{6h}} \left(\delta^{\alpha \beta} \sum_{\gamma \delta}  s_i^{\gamma} s_i^{\delta} - 1 \right)\;,
\end{split}
\end{equation}
for $\bm{a}$ the 12 nearest-neighbour vectors. Therefore, we can compute the physical eigenvalues of the Hessian by a final rotation in the tangent space. At the minimum, for example $\bm{s}_i=(1,1,1)^T/\sqrt{3}$, we obtain
\begin{equation}\label{eq:spectrumMin}
    \varepsilon_{\bm{q},\mu}^{(m)}= \varepsilon_{\bm{q}}+\Delta\varepsilon_\mu^{(m)},
\end{equation}
with the ferromagnetic exchange contribution:
\begin{equation}
    \varepsilon_{\bm{q}}=-2J\sum_{\bm{a}: |\bm{a}|>0} \left[\cos(\bm{q}\cdot\bm{a})-1\right]\;, 
\end{equation}
and the anisotropy contribution:
\begin{equation}
    \Delta\varepsilon_\mu^{(m)}=\frac{4}{3} K_{O_{h}} + 2R K_{D_{6h}}\;.
\end{equation}
Similarly at the saddle point, for example for $\bm{s}_i=(1,-1,0)^T/\sqrt{2}$, we obtain
\begin{equation}\label{eq:spectrumSP}
    \varepsilon_{\bm{q},\mu}^{(s)}= \varepsilon_{\bm{q}}+\Delta\varepsilon_\mu^{(s)},
\end{equation}
with the anisotropy contribution:
\begin{multline}
    \Delta\varepsilon_\mu^{(s)}=\frac{1}{2} K_{O_{h}} - R K_{D_{6h}}\\+(-1)^\mu \sqrt{\frac{9}{4}K_{O_h}^2-RK_{O_h}K_{D_{6h}}+R^2K_{D_{6h}}^2}\;.
\end{multline}
As $\alpha=0.01$ for Cobalt~\cite{Weber2019}, we have $ \alpha \ll \sqrt{4|\varepsilon_{\bm{q},1}^{\{\bm{s}_i\}}/\varepsilon_{\bm{q},2}^{\{\bm{s}_i\}}|} \sim 1$ $\forall \bm{q}$ and we can simply rewrite the prefactor of the transition rate as in Ref.~\cite{prb}: 
\begin{equation}
    \nu=\frac{3w_{\bm{0}}^{(m)}}{\pi}\prod_{\bm{q}\neq \bm{0}} \frac{w_{\bm{q}}^{(m)}}{w_{\bm{q}}^{(s)}}\;,
\end{equation}
where $w_{\bm{q}}^{(m)}$ and $w_{\bm{q}}^{(s)}$ are the positive frequencies of the spin wave modes solving Eqs.~(\ref{eq:dynamicalMode_sp}) and~(\ref{eq:dynamicalMode_min}) in the underdamped limit ($\alpha$=0) for which the spin waves are degenerate, i.e.
\begin{equation}
   \hbar w_{\bm{q}}^{\{\bm{s}_i\}}=\sqrt{\varepsilon_{\bm{q},1}^{\{\bm{s}_i\}}\varepsilon_{\bm{q},2}^{\{\bm{s}_i\}}}\;.
\end{equation}
Using the decomposition of Eqs.~(\ref{eq:spectrumMin}) and~(\ref{eq:spectrumSP}) and by expanding in terms of $(\Delta \varepsilon_1^{\{\bm{s}_i\}}+\Delta \varepsilon_2^{\{\bm{s}_i\}})/\varepsilon_{\bm{q}}$, we obtain
\begin{equation}
      \hbar  w_{\bm{q}}^{\{\bm{s}_i\}} \approx \hbar w_{\bm{q}} + \mathcal{C}^{\{\bm{s}_i\}}\;,
\end{equation}
with the exchange spin wave spectrum $\hbar w_{\bm{q}}=\varepsilon_{\bm{q}}$ plotted in Fig.~\ref{fig:spectrum} and the mean curvature of the anisotropy energy at the stationary states:
\begin{equation}
    \mathcal{C}^{\{\bm{s}_i\}}=\frac{\Delta \varepsilon_1^{\{\bm{s}_i\}}+\Delta \varepsilon_2^{\{\bm{s}_i\}}}{2}\;.
\end{equation}

\begin{figure}[h]
\includegraphics[width=1.0\linewidth,trim=0cm 0cm 0cm 0cm, clip]{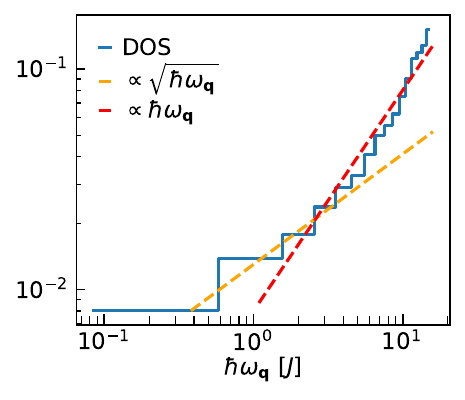}
\caption{\label{fig:dos} Density of states (DOS) of the exchange spin wave spectrum on an FCC particle with a diameter of 5nm. The square root dependence at low energy is typical for the quadratic dispersion near the $\Gamma$ point in Fig.~\ref{fig:spectrum}. At higher energy, the density of states depends on the lattice. A linear trend fits best the bulk of the density of states.} 
\end{figure}

We remark therefore that the spin wave spectrum is constantly gaped by the mean curvature of the anisotropy energy at the stationary states.
We expand now in the same parameter the product in the prefactor:
\begin{equation}
    \prod_{\bm{q}\neq \bm{0}} \frac{w_{\bm{q}}^{(m)}}{w_{\bm{q}}^{(s)}}\approx  \exp\left[\left(\mathcal{C}^{(m)}-\mathcal{C}^{(s)}\right)\sum_{\bm{q}\neq \bm{0}} \frac{1}{w_{\bm{q}}} \right]\;.
\end{equation}
According to Fig.~\ref{fig:dos}, the density of states of the exchange spin wave spectrum is essentially linear making possible to write
\begin{equation}
 \prod_{\bm{q}\neq \bm{0}} \frac{w_{\bm{q}}^{(m)}}{w_{\bm{q}}^{(s)}}\approx  \exp\left[\left(\mathcal{C}^{(m)}-\mathcal{C}^{(s)}\right) \frac{2N}{\max(\varepsilon_{\bm{q}})} \right]\;,
\end{equation}
with $\max(\varepsilon_{\bm{q}})=17J$ from the numerics. 
Finally, the $\bm{q}=\bm{0}$ contribution to the prefactor can be rewritten to give
\begin{equation}\label{eq:prefactorQ0}
    \frac{3w_{\bm{0}}^{(m)}}{\pi}=\frac{3\mathcal{C}^{(m)}}{\pi \hbar}\;.
\end{equation}

Bringing everything together, we obtain explicitly:
\begin{equation}\label{eq:analyticTransitionRate}
   \Gamma=  \nu \exp\left[-\frac{N}{K_BT}\left(\frac{K_{O_h}}{12}+RK_{D_{6h}}\right)\right]\;,
\end{equation}
with the prefactor
\begin{multline}\label{eq:analyticPrefactor}
        \nu=\frac{3}{\pi \hbar}\left(\frac{4}{3}K_{O_h}+2RK_{D_{6h}}\right) \\ \times \exp\left[\frac{N}{17J}\left(\frac{5}{3}K_{O_h}+6RK_{D_{6h}}\right)\right]\;.
\end{multline}

\bibliography{bibfileBocquet}

\end{document}